\documentclass[12pt]{article}
\usepackage{amsmath}
\usepackage{amsfonts}
\usepackage{amssymb}
\usepackage{graphicx,color}
\usepackage{color}
\usepackage{young}
\usepackage[vcentermath]{youngtab}
\usepackage{cite}
\usepackage[utf8]{inputenc}
\usepackage{slashed}
\usepackage{array}
\usepackage{graphicx}
\usepackage{ulem}

\usepackage{soul,xcolor}

\newcommand{\de}{\delta}
\newcommand{\ep}{\epsilon}

\newcommand{\be}{\begin{equation}}
\newcommand{\ee}{\end{equation}}
\newcommand{\bea}{\begin{eqnarray}}
\newcommand{\eea}{\end{eqnarray}}

\begin{document}

\thispagestyle{empty}

\setcounter{page}{0}

\mbox{}
\vspace{-35mm}

\begin{center} {\bf \Large  Supersymmetric electric-magnetic duality of hypergravity} 

\vspace{1.1cm}

Claudio Bunster$^{1,2}$,  Marc Henneaux$^{1,3}$,  Sergio H\"ortner$^3$, Amaury Leonard$^{3}$

\footnotesize
\vspace{.4 cm}

${}^1${\it Centro de Estudios Cient\'{\i}ficos (CECs), Casilla 1469, Valdivia, Chile}

\vspace{.1cm}

${}^2${\it Universidad Andr\'es Bello, Av. Rep\'ublica 440, Santiago, Chile}

\vspace{.1cm}

${}^3${\it Universit\'e Libre de Bruxelles and International Solvay Institutes, ULB-Campus Plaine CP231, B-1050 Brussels, Belgium} \\

\vspace {9mm}

\end{center}
\centerline{\bf Abstract}
\vspace{.5cm}

Hypergravity is the theory in which the graviton, of spin-$2$, has a supersymmetric partner of spin-$5/2$. There are ``no-go'' theorems that prevent interactions in these higher spin theories. However, it appears that one can circumvent them by bringing in an infinite tower of higher spin fields. With this possibility in mind, we study herein the electric-magnetic duality invariance of hypergravity. The analysis is carried out in detail for the free theory of the spin-$(2,5/2)$ multiplet, and it is indicated how it may be extended to the infinite tower of higher spins. Interactions are not considered. The procedure is the same that was employed recently for the spin-$(3/2,2)$ multiplet of supergravity. One introduces new potentials (``prepotentials'') by solving the constraints of the Hamiltonian formulation. In terms of the prepotentials, the action is written in a form in which its electric-magnetic duality invariance is manifest. The prepotential action is local, but the spacetime invariance is not manifest. Just as for the spin-$2$ and spin-$(3/2,2)$ cases, the gauge symmetries of the prepotential action take a form similar to those of the free conformal theory of the same multiplet. The automatic emergence of gauge conformal invariance out of demand of manifest duality invariance, is yet another evidence of the subtle interplay between duality invariance and spacetime symmetry. We also compare and contrast the formulation with that of the analogous spin-$(1,3/2)$ multiplet.

\vspace{.8cm}
\noindent

\newpage


\section{Introduction}
\setcounter{equation}{0}

There are two possible supersymmetric partners to the graviton: a massless spin-3/2 particle or a massless spin-5/2 particle. The first possibility leads to supergravity and has undergone spectacular developments. The second possibility has been called ``hypergravity" \cite{Aragone:1979hx}. It was explored in the early days of supergravity as a potentially interesting alternative \cite{Aragone:1979hx,Berends:1979wu,Berends:1979kg}.

However, with the no-go theorems of \cite{Aragone:1979hx,Aragone:1980rk} preventing consistent couplings between the massless spin-two and spin-five-half fields, the study of this other supersymmetric extension of the Pauli-Fierz theory was abandoned, except in three spacetime dimensions where a consistent interacting theory exists \cite{Aragone:1983sz}.  

But the work of the Lebedev school \cite{Fradkin:1986ka,Fradkin:1986qy,Fradkin:1987ks,Vasiliev:1990en,Vasiliev:1999ba} has changed the picture since it has been proved to be possible to overcome the spin-two barrier identified by the no-go theorems by considering an infinite number of fields (for recent reviews, see \cite{Bekaert:2010hw,Didenko:2014dwa}).  It is therefore of interest to re-consider hypergravity.  This is the purpose of the present paper, which deals with the electric-magnetic duality properties of the combined spin-two/spin-five-half system. 

It was shown long ago \cite{Deser:1976iy} that one can reformulate Maxwell theory in a way that makes the duality invariance of the action (and not just of the equations of motion) manifest.   This is achieved by solving the constraint (Gauss' law) of the Hamiltonian formulation through the introduction of a  vector potential  for the electric field. Duality is then a rotation in the internal plane of the ordinary vector potential  for the magnetic field and the new potential for the electric field.  This approach to electric-magnetic duality has been successfully extended to various systems \cite{Henneaux:1988gg,ScSe,Deser:1997mz,Hillmann:2009zf,Bunster:2011aw,Bunster:2011qp}, including linearized gravity in four  \cite{Henneaux:2004jw,Bunster:2012km} and higher \cite{Bunster:2013oaa} dimensions.

We provide in this work the supersymmetric extension of  the manifestly electric-magnetic duality invariant spin-two action of \cite{Henneaux:2004jw} obtained by supersymmetrizing along the ``hypergravity route".   Since we only consider the spin-(2,5/2)-system, we are forced to limit ourselves to the free theory.  Comments on interactions, which need even higher spin fields, are made in the conclusions.

In order to solve the constraints of the Hamiltonian formalism, we introduce a prepotential for the spin-5/2 field that is the supersymmetric partner of the prepotentials for the spin-2 field.  This is in close analogy to what was done for standard supergravity in \cite{Bunster:2012jp}, but the constraints are now more intricate.  The spin-5/2 prepotential is a symmetric tensor-spinor, and the original spin-5/2 field depends on its second derivatives, while the spin-3/2 field depended only on the first derivatives of its prepotential.   We also find again, just as in \cite{Bunster:2012jp}, that  in order for electric-magnetic duality to commute with supersymmetry it is necessary to define the former as acting chirally on the spinors. (The interplay between chirality and duality invariance was already stressed in \cite{Deser:1977ur}.)

Our work is organized as follows.  In the next section, we recall the covariant formulation of  hypergravity.  We then turn in Section \ref{HamiltonianForm} to the Hamiltonian formulation of the theory as this is necessary to manifestly exhibit duality.  Section \ref{Solving} provides the solution of the constraints of the Hamiltonian formulation in terms of the prepotentials.  In Section \ref{SUSYPrep}, we write the supersymmetry transformations for the prepotentials. In Section \ref{FormOfSusyAction}, we give the expression of the action in terms of the prepotentials. Finally,  Section \ref{Conclusions} is devoted to conclusions and prospects.  The more technical derivations are relegated to Appendix \ref{Technical}.  Appendix \ref{132Mult} pursues the similar duality analysis for the spin-$(1, 3/2)$ system.

\section{Free theory in covariant form}
\label{FreeLS}
\setcounter{equation}{0}

\subsection{Action}
The action describing the combined system of a free spin-2 massless field $h_{\mu \nu} = h_{\nu \mu}$ with a free spin-5/2 massless field $\psi_{\mu \nu} = \psi_{\nu \mu}$ is the sum of the Pauli-Fierz action and the spin 5/2 action \cite{Rarita:1941mf,Schwinger:1970xc,Singh:1974rc,Fang:1978wz}, 
\begin{equation}
S[h_{\mu \nu}, \psi_{\mu \nu}]=\int d^{D}x \left({\mathcal L}_2 + {\mathcal L}_{\frac52} \right) \label{Action0}
\end{equation}
where
\begin{equation}
{\mathcal L}_2=-\frac{1}{4}\left(\partial^{\rho}h^{\mu\nu}\partial_{\rho}h_{\mu\nu}-2\partial_{\mu}h^{\mu\nu}\partial_{\rho}h^{\rho}_{\ \nu}+2\partial^{\mu}h\partial^{\nu}h_{\mu\nu}-\partial^{\mu}h\partial_{\mu}h\right),
\end{equation}
and
\begin{equation}
{\mathcal L}_{\frac52} = 
i   \left(
-  \frac{1}{2} \ \bar{\psi}^{\mu\nu} \slashed{\partial} \psi_{\mu\nu}
+  2 \ \bar{\psi}^{\mu\rho}\gamma_\rho \partial^{\nu} \psi_{\mu\nu}
 -  \bar{\psi}^{\mu \rho}\gamma_\rho \slashed{\partial} \slashed{\psi}_{\mu}
 -  \bar{\psi}^{\mu \rho}\gamma_\rho\partial_{\mu} \psi
 +  \frac{1}{4} \ \bar{\psi}\slashed{\partial}\psi  
\right) .   \label{action_lag_5_2}
\end{equation}

Our convention are as follows.
We work in Minkoswski spacetime with  signature $- \ + \ + \ +$. Greek indices take values from $0$ to $3$ while Latin indices run from $1$ to $3$.  The covariant trace is $h\equiv h^{\mu}_{\phantom{\mu}\mu}$. The spin $5/2$ field is a symmetric Majorana spinor-tensor $\psi_{\mu\nu}$. We note  $\psi \equiv \psi^{\mu}_{\phantom{\mu}\mu}$,  $\slashed{A} \equiv \gamma^\mu A_\mu$, $\slashed{\psi}_{\mu}\equiv \gamma^\nu \psi_{\mu \nu}$. 

The Dirac matrices are in a Majorana representation (matrices $\gamma_{\mu}$ real, $\gamma_0^T = - \gamma_0$ and $\gamma_k^T = \gamma_k$ where $T$ denotes the transposition; and so $\gamma^{\mu \dagger} = \gamma^{\mu T} =\gamma^{0} \gamma^{\mu} \gamma^{0}$). In addition, $\gamma_5 \equiv  \gamma_0 \gamma_1 \gamma_2 \gamma_3 = - 1/4! \ \epsilon^{\mu \nu \rho \sigma} \gamma_{\mu} \gamma_{\nu} \gamma_{\rho} \gamma_{\sigma}$ (where $\epsilon^{0123} = - 1 =  \epsilon_{0123}$), which implies $\gamma_5^{\dagger} = -\gamma_5$, $\gamma_5^{T} = - \gamma_5$ et $\left( \gamma_5 \right)^2 =- I$. Finally, $\gamma^{\mu\nu} \equiv \frac{1}{2} \left[ \gamma^{\mu} , \gamma^{\nu} \right] = \gamma^{[\mu}\gamma^{\nu]}$ (and $\delta^{\mu\nu\rho}_{\alpha\beta\gamma} \equiv 6 \ \delta^{[\mu}_{\alpha} \delta^{\nu}_{\beta}\delta^{\rho]}_{\gamma}$), etc.

\subsection{Gauge symmetries}

The action (\ref{Action0}) is invariant under linearized diffeomorphisms
\begin{equation}
\delta_{\text{gauge}} h_{\mu\nu} = \partial_{\mu} \xi_{\nu} \ + \ \partial_{\nu} \xi_{\mu} .
\end{equation}
and gauge transformations of the spin-$5/2$ field,
\begin{equation}
\delta_{\text{gauge}} \psi_{\mu\nu} = \partial_{\mu} \zeta_{\nu} \ + \ \partial_{\nu} \zeta_{\mu} ,
\end{equation}
where the ``spin-$3/2$" gauge parameter $\zeta_\mu$ is $\gamma$-traceless,
\begin{equation}\slashed{\zeta} = 0 .
\end{equation}
Because of the tracelessness condition, one can express $\zeta_0$ in terms of $\zeta_k$,
\begin{equation}
\zeta_0 = - \gamma_0 \gamma^k \zeta_k.
\end{equation}
There are three independent fermionic gauge symmetries.

\subsection{Supersymmetry}
The action is also invariant under rigid supersymmetry, which reads,
\begin{eqnarray}
\delta_{\text{SUSY}} h_{\mu\nu} &=& 8i \ \bar{\epsilon}\psi_{\mu\nu}, \label{susy_h}
\\
\delta_{\text{SUSY}} \psi_{\mu\nu} &=& 
\left( \partial_{\mu} h_{\nu\rho} \ + \ \partial_{\nu} h_{\mu\rho} \right) \gamma^{\rho} \epsilon
\ - \ 2 \ \partial_{\rho} h_{\mu\nu} \gamma^{\rho} \epsilon
\nonumber \\ &&
 +  \ \left( \epsilon_{\mu \lambda \sigma \rho} \partial^{\lambda} h_{\nu}^{\phantom{\nu} \sigma} 
\ + \ \epsilon_{\nu \lambda \sigma \rho} \partial^{\lambda} h_{\mu}^{\phantom{\nu} \sigma} \right)
\gamma^{\rho} \gamma_5 \epsilon
. \qquad \label{susy_psi}
\end{eqnarray}
The supersymmetry parameter $\epsilon$ is a constant spinor.

\subsection{Equations of motion}
The equations of motion can be written as
\begin{eqnarray}
&&R_{\mu \nu}= 0  \label{Einstein}\\
&& \gamma^\rho \Omega_{\mu \nu \rho} \label{EOM52}= 0
\end{eqnarray}
where $R_{\mu \nu}$ is the linearized Ricci tensor and $\Omega_{\mu \nu \rho}$ is given by \cite{Berends:1979wu}
\begin{equation}
\Omega_{\mu \nu \rho} = \frac12(\partial_{\mu} \psi_{\nu \rho} +\partial_{\nu} \psi_{\mu \rho} -\partial_{\rho} \psi_{\mu \nu}).
\end{equation}
The equations of motion and the action are invariant under electric-magnetic duality rotations in the internal two-dimensional space spanned the Riemann tensor and its dual.  They are also invariant under chirality rotations of the spinor fields,
\begin{equation}
\delta_{\text{chiral}} h_{\mu \nu} = 0, \; \; \; \; \delta_{\text{chiral}}  \psi_{\mu \nu} =  \lambda \gamma_5 \psi_{\mu \nu}
\end{equation}

That the action is invariant under chirality rotations is manifest. To display explicitly its electric-magnetic duality invariance, one must go to the first-order formalism and introduce prepotentials for the spin-two field. Supersymmetry forces one to then introduce  prepotentials for the spin-five-half partner, as we now discuss.  We will also see that a definite chirality rotation must accompany a duality transformation if duality is to commute with supersymmetry.


\section{Hamiltonian form}
 \setcounter{equation}{0}
 \label{HamiltonianForm}
 
 \subsection{Spin-2 part}
 As it is well known, the Paul-Fierz action takes the canonical form 
 \begin{equation}
S_2[h_{mn}, \pi^{mn}, n, n_m] =\int dt d^{3}x\left[\pi^{mn}\dot{h}_{mn}-{\cal{H}}- nC -n_m C^m \right]
\end{equation}
where ${\cal{H}}$ is the Hamiltonian density:
\begin{eqnarray}
{\cal{H}}&=& \pi_{mn}\pi^{mn}-\frac{\pi^{2}}{2}+\frac{1}{4}\partial_{r}h_{mn}\partial^{r}h^{mn}-\frac{1}{2}\partial_{m}h^{mn}\partial_{r}h^{r}_{\ n}+\nonumber \\
&& \hspace{2cm} + \frac{1}{2}\partial^{m}h\partial^{n}h_{mn}-\frac{1}{4}\partial_{m}h\partial^{m}h.
\end{eqnarray}
The components $h_{0m}\equiv n_m$ and $h_{00}\equiv 2 n$ only appear linearly and multiplied by terms with no time derivatives, and are thus Lagrange multipliers for the constraints
\begin{equation}
C^m \equiv  -2 \partial_{n}\pi^{mn}=0 \ \ \ \ \ C \equiv  -\Delta h+\partial_{m}\partial_{n}h^{mn}=0.
\label{Spin2Constraints} \end{equation}

The constraints generate the gauge symmetries (linearized diffeomorphisms) through the Poisson brackets.  These are, in terms of the canonical variables,
\begin{equation}
\pi^{mn}\rightarrow \pi^{mn}-\partial^{m}\partial^{n}\xi^{0}+\delta^{mn}\xi^{0} \label{b}
\end{equation}
\begin{equation}
h_{mn}\rightarrow h_{mn}+\partial_{m}\xi_{n}+\partial_{n}\xi_{m} \label{aa}
\end{equation}

 \subsection{Spin-5/2 part}
 We turn now to the spin-5/2 action.  Its Hamiltonian formulation has been performed in \cite{Aragone:1979hw,Borde:1981gh} but our treatment differs from these earlier works in that we do not fix the gauge at any stage and solve the (first-class) constraints through the introduction of  prepotentials on which the fields depend locally. 
 
The action for the spin-five-half field, being already of first order, is almost in canonical form.  Because there are gauge symmetries, some of the components of $\psi_{\mu \nu}$ are Lagrange multipliers for the corresponding first class constraints, while the other components define the phase space of the system.  What differentiates the Lagrange multipliers  from the standard phase space variables is that the gauge variations of the former contain the time derivatives $\dot{\zeta}_m$ of the gauge parameters $\zeta_m$, while the gauge variations of the latter do not involve $\dot{\zeta}_m$.  
 
To bring the action for the spin-five-half field to canonical form, we make the redefinition
\begin{eqnarray} 
&&\Xi = \psi_{00} - 2\ \gamma^{0} \gamma^{k} \psi_{0k} , \label{IV_variablechange}\\
&&\psi_{0k} = \psi_{0k}, \\
&&\psi_{mn} = \psi_{mn}.
\end{eqnarray}
One then has the following transformation rules,
\begin{eqnarray}
\delta_{\text{gauge}}\Xi &=& - \ 2\ \gamma^{k} \gamma^{l} \partial_{k} \zeta_{l} \label{IV_gaugetransf_00'} \\
\delta_{\text{gauge}}\psi_{0k} &=& \partial_{0} \zeta_{k} + \ \gamma^{0} \gamma^{l} \partial_{k} \zeta_{l} \label{IV_gaugetransf_0k}\\
\delta_{\text{gauge}} \psi_{kl} &=& \partial_{k}\zeta_{l} + \partial_{l}\zeta_{k}, \label{IV_gaugetransf_mn}
\end{eqnarray}
from which one immediately identifies the $\psi_{0k}$ components as the Lagrange multipliers.  

Indeed, the action takes the form,
\begin{eqnarray}
S_{\frac52}[\Xi, \psi_{mn}, \psi_{0k}] &=& \int d^{4} x  \lbrace  \Theta^{A} ( \psi_{B} ) \dot{\psi}_{A} 
 +  \psi_{0k}^T \mathcal{F}_{k}\left( \psi_{B} \right) 
 -  \mathcal{H}\left( \psi_{B} \right)  \rbrace, \hspace{.5cm}
\label{IV_action_newform}
\end{eqnarray}
with $(\psi_A) \equiv (\Xi, \psi_{mn})$ and were the Hamiltonian $\mathcal{H}$,  the constraint functions $\mathcal{F}_k$ and the symplectic potential $\Theta^A$ are explicitly given by,
\begin{eqnarray}
\mathcal{H} &=& 
- \ \frac{3i}{4} \ \bar{\Xi} \gamma^{k} \partial_{k} \Xi
\ + \ \frac{i}{2} \ \bar{\psi}^{kl} \gamma^{m} \partial_{m} \psi_{kl}
\ - \ 2i\ \bar{\psi}^{kl} \gamma_k \partial^{m} \psi_{lm} 
\ - \ i \ \bar{\psi}^{kl} \gamma_{k} \partial_{l} \Xi
\nonumber \\ &&
 + \ i \ \bar{\psi}^{kl} \gamma_{k} \partial_{l} \psi
\ + \ \frac{i}{2} \ \bar{\Xi} \gamma^{k} \partial_{k} \psi
\ - \ \frac{i}{4} \ \bar{\psi} \gamma^{l} \partial_{l}\psi
\ + \ i \ \bar{\psi}_{kl} \gamma^{l} \gamma^{m} \gamma^{n} \partial_{m} \psi^k_{\ n}
, \qquad \qquad \label{IV_hamiltonian} 
\\
\mathcal{F}_k &=&
- \ i \ \gamma^0 \left[
\partial_{k} \Xi
\ - \ 2 \ \partial^{l} \psi_{kl}
\ + \ \partial_{k} \psi
\ + \ \gamma_{k} \gamma^{l} \partial_{l} \Xi
\ + \ 2 \ \gamma^{l} \gamma^{m} \partial_{l} \psi_{km}
\ - \ \gamma_{k} \gamma^{l} \partial_{l} \psi \right.
\nonumber \\ &&
\left. + \ 2 \ \gamma_k \gamma^l \partial^m \psi_{lm} \right] , \qquad \qquad \label{IV_constraints}
\\
\Theta_{\Xi} &=&
 \frac{i}{4}\ \Xi^{T}
\ - \ \frac{i}{2} \ \psi^{T}, \qquad \qquad \label{IV_theta_00}
\\
\Theta_{kl} &=&
\frac{i}{2} \ \psi^{T}_{kl}
\ - \ \frac{i}{4} \ \psi^{T} \delta_{kl}
\ - \ \frac{i}{2} \ \psi^{T}_{km} \gamma^{m} \gamma_{l}
\ - \ \frac{i}{2} \ \psi^T_{lm} \gamma^{m} \gamma_{k}. \qquad \qquad \label{IV_theta_kl}
\end{eqnarray}
Here and from now on, $\psi$ denotes the {\it spatial} trace $\psi = \psi^k_{\; \; k}$.

The action (\ref{IV_action_newform}) is the searched-for action in canonical form because the symplectic form $\sigma = d \Theta$ is non-degenerate and yields the following brackets between the canonical variables,
\begin{eqnarray}
\left\lbrace \Xi \left( \vec{x} \right)  , \Xi  \left( \vec{x}' \right) \right\rbrace_D &=& 
- \ \frac{5i}{4} \ \delta \left(  \vec{x} - \vec{x}' \right) ,\label{IV_diracbracket_0000}
\\
\left\lbrace \Xi\left( \vec{x} \right)  , \psi_{kl}  \left( \vec{x}' \right) \right\rbrace_D &=& 
\frac{i}{4} \ \delta_{kl} \ \delta \left(  \vec{x} - \vec{x}' \right) , \label{IV_diracbracket_00kl}
\\
\left\lbrace \psi_{kl} \left( \vec{x} \right)  , \psi_{mn}  \left( \vec{x}' \right) \right\rbrace_D &=& 
\left[ - \ \frac{i}{4} \ ( \delta_{km} \delta_{ln} + \delta_{kn} \delta_{lm} ) 
\ + \ \frac{i}{4} \ \delta_{kl} \delta_{mn} \right.
\nonumber \\ &&
 + \left. \ \frac{i}{8} \ \left( \delta_{km} \gamma_{ln} + \delta_{kn} \gamma_{lm} + \delta_{lm} \gamma_{kn} + \delta_{ln} \gamma_{km} \right) \right] \ \delta \left(  \vec{x} - \vec{x}' \right), \hspace{1.2cm}
 \label{IV_diracbracket_klmn}
\end{eqnarray}
These brackets would appear as Dirac brackets had one introduced conjugate momenta for the fermionic variables $\psi_A$ and eliminated the corresponding second class constraints that express these momenta in terms of the $\psi_A$ through the Dirac bracket procedure - hence the notation $\lbrace, \rbrace_D$.

The variables $\psi_{0k}$ are the Lagrange multipliers for the first-class constraints $\mathcal{F}_k= 0$, which are easily verified to generate the fermionic gauge transformations (\ref{IV_gaugetransf_00'}) and (\ref{IV_gaugetransf_mn}) through the Dirac bracket.


\section{Solving the constraints - Prepotentials}
\label{Solving}
\setcounter{equation}{0}
\subsection{Bosonic constraint}
In order to exhibit gravitational duality, it is necessary to solve the spin-2 constraints (\ref{Spin2Constraints}) through the introduction of prepotentials  \cite{Henneaux:2004jw}.  The momentum constraints $C^m = 0$ are solved as
\begin{equation}
\pi^{kl} =
\epsilon^{kpr}\epsilon^{lqs} \partial_p \partial_q P_{rs} , 
\end{equation}
while the Hamiltonian constraint $C=0$ yields
\begin{equation}
h_{kl} =
\epsilon_{kmn} \partial^m \Phi^n_{\phantom{n}l} \ + \ \epsilon_{lmn} \partial^m \Phi^n_{\phantom{n}k} + \partial_k u_l + \partial_l u_k.
\end{equation}
Here, the prepotentials $P_{mn}$ and $\Phi_{mn}$ are both symmetric.  The prepotential $u_k$ drops out by gauge invariance from the action and is usually set equal to zero for that reason.

The gauge symmetries of the prepotentials take the remarkable form of those of linearized conformal gravity,
\begin{equation}
\delta Z^a_{ij} = \partial_i \xi^a_{j} + \partial_j \xi^a_{i} + 2 \epsilon^a \delta_{ij} \label{GaugeZ}
\end{equation}
where $(Z^a_{ij}) \equiv (P_{ij}, \Phi_{ij})$ ($a= 1,2$).  These transformations take into account the redundancy present in the definition of the prepotentials.   In the case of $P_{ij}$,  the transformation (\ref{GaugeZ}) also contains the original gauge symmetries (\ref{b}) of  $\pi^{kl}$ generated by the Hamiltonian constraint.    In the case of $\Phi_{ij}$, the transformation (\ref{GaugeZ}) induces spatial diffeomorphisms (\ref{aa}) that are divergence-free (i.e., equal to the curl of a vector field, $\xi^m = \epsilon^{mpq} \partial_p \zeta_q$).  The most general diffeomorphism is accounted for by arbitrary shifts $u_k \rightarrow u_k + \xi_k$ of the prepotential $u_k$.

Because of the gauge symmetries (\ref{GaugeZ}), each unconstrained prepotential contains $6$ (components) minus $3$ (diffeomorphism gauge parameters) minus $1$ (Weyl rescaling gauge parameter) = $2$ independent, physical functions. This gives $2 \times 2 = 4$ independent, physical functions, corresponding to the four independent initial data needed to describe the two physical helicities of the massless spin-$2$ field.

\subsection{Fermionic constraint}
The fermionic constraint (\ref{IV_constraints}) can be rewritten in the simpler form
\begin{equation}
\partial_k \Xi + \partial_k \psi + 2\gamma^{ij} \partial_i \psi_{kj} = 0 \label{C5/2}
\end{equation}
Indeed, if one multiplies (\ref{C5/2}) with $k$ replaced by $l$ by the invertible operator $- i \gamma^0 \left(\delta^l_k + \gamma_k \gamma^l \right)$, one gets ${\mathcal F}_k = 0$.

The equation (\ref{C5/2}) itself can be rewritten as $\partial_i \left(\delta^i_k \Xi + \delta^i_k \psi + 2 \gamma^{ij} \psi_{kj} \right) = 0$, which by virtue of the Poincar\'e lemma implies
\begin{equation}\delta^i_k \Xi + \delta^i_k \psi + 2 \gamma^{ij} \psi_{kj} = \partial_r A^{[ir]}_{\; \; \; \; k} \label{KeyEq0} \end{equation}
for some tensor $A^{[ir]}_{\; \; \; \; k} = - A^{[ri]}_{\; \; \; \; k}$, which can be decomposed into irreducible components as 
$$ A^{[ir]}_{\; \; \; \; k} = \delta^i_k a^r - \delta^r_k a^i + \epsilon^{irm} m_{mk} $$ with $m_{mk} = m_{km}$.  The equation (\ref{KeyEq0}) can be solved to yield  $\psi_{ij}$ and $\Xi$  in terms of $a^k$ and $m_{ij}$ as
\begin{eqnarray}
&& 2 \psi_{lk} = - \partial_k a^l + \frac13 \delta_{lk} \partial_q a^q + \epsilon_l^{\; \; qs} \partial_q m_{sk} \nonumber \\
&& \hspace{1cm} + \frac12 \gamma_l \gamma_m \left(\partial_k a^m - \frac13 \delta^m_k \partial_q a^q + \epsilon^{qms} \partial_q m_{sk} \right) \label{psiam}\\
&& \Xi = - \frac23 \partial_q a^q - \psi \label{Xiam}
\end{eqnarray}
Now, the right-hand side of (\ref{psiam}) is not symmetric in general, while $\psi_{lk}$ is symmetric.  The system of equations (\ref{KeyEq0}) is overdetermined.  A symmetric solution of (\ref{KeyEq0}) exists if and only if $A^{[ir]}_{\; \; \; \; k}$ -- or equivalently, $a^q$ and $m_{ij}$ --  fulfills the constraints expressing that  $\psi_{lk} = \psi_{kl}$.  This condition is a differential constraint which can be rewritten, by virtue of the Poincar\'e lemma, as
\begin{eqnarray}
&& a_k \delta^q_l - a_l \delta^q_k + \frac12 \left(\gamma_l \gamma_m \delta^q_k - \gamma_k\gamma_m \delta^q_l \right) a^m - \frac13 \gamma_{lk} a^q \nonumber \\
&& \; \; \; + \epsilon_l^{\; \; qs} m_{sk}  - \epsilon_k^{\; \; qs} m_{sl} + \frac12 \gamma_l \gamma_m \epsilon^{qms} m_{sk} - \frac12 \gamma_k \gamma_m \epsilon^{qms} m_{sl} = \partial_p B_{lk}^{qp} \hspace{1cm} \label{KeyEq1}
\end{eqnarray}
for some tensor $B_{lk}^{qp}$ antisymmetric in $l,k$ and $q,p$,  $B_{lk}^{qp} = -B_{kl}^{qp} = -B_{lk}^{pq}$.  Trading each antisymmetric pair for one single index using the Levi-Civita tensor, this tensor $B_{lk}^{qp}$ is equivalent to a tensor $B_{ij}$, $B^{lkqp} \sim \epsilon^{lki} \epsilon^{qpj} B_{ij}$, which can be decomposed into a symmetric part $\Sigma_{ij}$ and an antisymmetric part $\epsilon_{ijk} V^k$.  

The system (\ref{KeyEq1}) is a system of 9 linear spinorial equations for the 9 spinorial unknows $a^q$ and $m_{ij} = m_{ji}$.  We leave it to the reader to verify that this system can be solved uniquely for  $a^q$ and $m_{ij}$ in terms of the first derivatives of $\Sigma_{ij}$ and $V^k$, which are therefore unconstrained.  Next, one substitutes the resulting expressions in (\ref{psiam}) and (\ref{Xiam}), allowing in addition for a general gauge transformation term in which one absorbs all gauge transformation terms involving  $\Sigma_{ij}$ and $V^k$ through redefinitions.  Then one finally gets the searched-for expressions
\begin{eqnarray}
&& \Xi = \triangle \Sigma + \partial^a \partial^b \Sigma_{ab}- 2 \gamma^k \gamma^l \partial_k v_l \label{XiSigma1}\\
&& \psi_{ij} = \delta_{ij} \triangle \Sigma  - \delta_{ij} \partial^a \partial^b \Sigma_{ab} - 2 \triangle \Sigma_{ij} \nonumber \\
&& \hspace{1.5cm} - \left( \gamma_{ia} \partial^a \partial^b \Sigma_{jb} + \gamma_{ja} \partial^a \partial^b \Sigma_{ib}
\right)  \nonumber \\
&& \hspace{1.5cm} + \left( \gamma_{ia} \triangle \Sigma^{a}_{\; \; j} + \gamma_{ja} \triangle \Sigma^{a}_{\; \; i}\right) \nonumber \\
&& \hspace{1.5cm} +\partial_i v_j + \partial_j v_i \label{psiSigma1}
\end{eqnarray}
of the fields $\Xi$ and $\psi_{ij}$ in terms of a spin 5/2 prepotential $\Sigma_{ij} = \Sigma_{ji}$  and a spin 3/2 prepotential $v_i$ (in which $V^k$ has been completely absorbed).

As a consistency check, it is easy to verify that these expressions identically fulfill the spinorial constraint (\ref{C5/2}), as they should. Note that the prepotential $v_i$ drops out from gauge-invariant expressions and plays for that reason a less fundamental role.  It is similar to the prepotential $u_i$ for the graviton.   

One can easily verify that the expressions are invariant under the gauge symmetries of the prepotential $\Sigma_{ij}$
\begin{equation}
\delta \Sigma_{ij} = \partial_i \mu_j + \partial_j \mu_i +\gamma_i \eta_j + \gamma_j \eta_i \label{TransSigma}
\end{equation}
where $\mu_i$ and $\eta_i$ are arbitrary. Indeed, one finds $\delta \Xi = 0$ and $\delta \psi_{ij} = 0$ provided one transforms simultaneously the other prepotential $v_i$ as
\begin{equation} \delta v_i = 2 \triangle \mu_i + \gamma_{ia} \partial^a \partial^b \mu_b - \gamma_{ia} \triangle \mu^a - \gamma^a \partial_a \eta_i + \gamma_i \partial^a \eta_a.\end{equation}
Note that the special choice $\eta_i = \gamma_i \phi$ yields the Weyl rescaling $2 \phi \delta_{ij}$ of the prepotential $\Sigma_{ij}$, which is therefore contained among the gauge symmetries.  The gauge transformations (\ref{TransSigma}) are redundant since one gets $\delta \Sigma_{ij} = 0$ for $\mu_i = \gamma_i \epsilon$ and $\eta_i = - \partial_i \epsilon$.  There are thus $3 \times 4$ (components of the vector-spinor $\mu_i$) plus $3 \times 4$ (components of the vector-spinor $\eta_i$) minus $1 \times 4$ (reducibility relations) $=  5 \times 4 = 20$ independant gauge parameters.

The gauge transformations of the prepotentials $v_i$ are mere shifts $ v_i \rightarrow v_i + \omega_i$ where $\omega_i$ are arbitrary vector-spinors.  These prepotentials contain therefore no degree of freedon and can be set for instance equal to zero.

The gauge transformations  (\ref{TransSigma}) are the only gauge symmetries of the prepotential $\Sigma_{ij}$, as can be seen by mere counting.  There are $6 \times 4 = 24$ components of the tensor-spinor $\Sigma_{ij}$.  These are unconstrained, but the gauge freedom removes as we have just seen $20$ components, leaving $4$ independent arbitrary functions needed to describe the physical helicities  of a massless spin-$5/2$ field. 

Another way to reach the same conclusion goes as follows.  One may rewrite (\ref{psiSigma1}) in terms of the Schouten tensor $S_{ij}[\Sigma]$ of $\Sigma_{ij}$,  
\begin{eqnarray}
&& S_{ij}[\Sigma] = \frac{1}{2} \left( \partial_i \partial^m \Sigma_{mj} + \partial_j \partial^m \Sigma_{mi}  - \triangle \Sigma_{ij} - \partial_i \partial_j \Sigma \right) \nonumber \\
&& \hspace{2cm} - \frac{1}{4} \left( \partial^m \partial^n \Sigma_{mn} - \triangle \Sigma \right) \delta_{ij}, \nonumber
\end{eqnarray}
as
\begin{equation}
\psi_{ij} =4 S_{ij} - 2 \gamma_i^{\;  a} S_{aj} - 2 \gamma_j^{\;  a} S_{ai} + \partial_i \bar{v}_j + \partial_j \bar{v}_i
\end{equation}
with
\begin{equation}
\bar{v}_i = v_i  - 2 \partial^m \Sigma_{mi} + \partial_i \Sigma + \gamma_i^{\; a} \partial^b \Sigma_{ab} - \gamma_i^{\; a} \partial_a \Sigma
\end{equation}
which transforms as
\begin{equation}
\delta \bar{v}_i =  \gamma ^k (\partial_i \eta_k - \partial_k \eta_i) + 3 \gamma_{i}^{\; \, km} \partial_m \eta_k 
\end{equation}
under (\ref{TransSigma}).  Similarly, one finds
\begin{equation}
\Xi = 4 S - 2 \gamma^k \gamma^l \partial_k \bar{v}_l.
\end{equation}
The Schouten tensor $S_{ij}[\Sigma]$ is invariant under linearized diffeomorphisms and ``sees" only the $\gamma$-transformations.  One finds explicitly 
\begin{eqnarray}
\delta S_{ij}[\Sigma] &=& \left( - \partial_i \partial_j + \frac12 \triangle \right) \gamma^k \eta_k \nonumber \\
&& \left( \frac12 (\gamma_i \partial_j + \gamma_j \partial_i) - \frac12 \delta_{ij} \gamma^m \partial_m \right) \partial^k \eta_k \nonumber \\
&& + \frac12 \gamma^k \partial_k \left( \partial_i \eta_j + \partial_j \eta_i \right) - \frac12 \left(\gamma_i \triangle \eta_j + \gamma_j \triangle \eta_i \right) \label{TransSchouten}
\end{eqnarray}
under (\ref{TransSigma}).

Now, if $\psi_{ij}$ and $\Xi$ are both equal to zero, one finds that the Schouten tensor is given by
\begin{equation}
8 S_{ij} = - 2 \partial^m\bar{v}_m  \delta_{ij}+ \partial_m(\gamma_j^{\; \, m} \bar{v}_i + \gamma_i^{\; \, m} \bar{v}_j) + \partial_i(\gamma_j^{\; \, m} \bar{v}_m +  \bar{v}_j) +\partial_j(\gamma_i^{\; \, m} \bar{v}_m +  \bar{v}_i) \label{FormOfS}
\end{equation}
with $\bar{v}_i$ constrained by
\begin{equation}
\gamma^k \gamma^l \partial_k \bar{v}_l +\partial^k \bar{v}_k = 0; \label{ConstraintV}
\end{equation}
By virtue of the Poincar\'e lemma, the general solution of (\ref{ConstraintV}) is given by 
\begin{equation}
\bar{v}_k = - 7 \gamma^m\partial_k \eta_m + 8 \gamma_k \partial^m \eta_m - 9 \gamma^m \partial_m\eta_k + 5 \gamma_k^{\; \, st} \partial_s \eta_t 
\end{equation} for some arbitrary $\eta_k$. 
But then, (\ref{FormOfS}) has exactly the form (\ref{TransSchouten}) of a $\gamma$-gauge transformation.  This means that one can set the Schouten tensor $S_{ij}[\Sigma]$ -- and hence also the Riemann tensor of $\Sigma_{ij}$ since the number of dimensions is three -- equal to zero by a $\gamma$-gauge transformation.  This implies that in that $\gamma$-gauge,  the tensor $\Sigma_{ij}$ itself is given by $\partial_i \mu_j + \partial_j \mu_i$,  or, in general, by (\ref{TransSigma}) if one does not impose any $\gamma$-gauge condition.  This shows that (\ref{TransSigma}) exhausts indeed the most general gauge freedom.

In parallel to what was found for the spin-2 prepotentials, the gauge symmetries of the spin-$\frac52$ prepotential take the same form as the gauge symmetries of the free conformal spin 5/2 field theory in 4 dimensions \cite{Fradkin:1985am}. To make the comparison it should be borne in mind that in \cite{Fradkin:1985am}, the redundancy in the gauge parameters is fixed by imposing the condition $\gamma^i \mu_i = 0$.

Of course, two physically different theories may have the same gauge symmetries. It is nevertheless quite remarkable that gauge conformal invariance should emerge automatically when one requires electric-magnetic duality invariance to be manifest. This is yet another evidence for the subtle interplay between duality invariance and spacetime symmetry.


\section{Supersymmetry transformations in terms of the prepotentials}
\label{SUSYPrep}
\setcounter{equation}{0}

A somewhat tedious but direct computation shows that the supersymmetry transformations of the bosonic prepotentials are given by
\begin{equation}
\delta\Phi_{ij}= -8i\bar{\epsilon}\chi_{ij} \label{SUSYPHI}
\end{equation}
and 
\begin{equation}
\delta P_{ij}=-8i  \bar{\epsilon} \gamma_5 \chi_{ij}. \label{SUSYP}
\end{equation}
Here, 
$$
\chi_{ij} = \frac{1}{2}(\epsilon_{jmn}\partial^{m}\Sigma_{i}^{\ n}+\epsilon_{imn}\partial^{m}\Sigma_{j}^{\ n}-\epsilon_{lmn}\gamma_{j}^{\ l}\partial^{m}\Sigma_{i}^{\ n}-\epsilon_{lmn}\gamma_{i}^{\ l}\partial^{m}\Sigma_{j}^{\ n})
$$ transforms under gauge transformations of $\Sigma_{ij}$ in the same way as the bosonic prepotentials, i.e.,
$$\delta \chi_{ij} = \partial_i \alpha_j + \partial_i \alpha_j  + \delta_{ij} \beta, $$  for some $\alpha_i$ and $\beta$. 
The details are given in  Appendix \ref{Technical}.   Similarly, one finds,
\begin{equation}
\delta \Sigma_{ij}=2\gamma_{5}\gamma_{0}(\Phi_{ij}-\gamma_{5}P_{ij})\epsilon , \label{SUSYS}
\end{equation}
where again the gauge transformations match.

Formulas (\ref{SUSYPHI}), (\ref{SUSYP}) and (\ref{SUSYS}) are the analogs for the spin-$(2,5/2)$ multiplet of the formulas (\ref{SusyPrepA10}), (\ref{SusyPrepA20}) and (\ref{SusyPrepChi0}) giving the supersymmetry transformations of the prepotentials of the spin-$(1, 3/2)$ multiplet.  Even though the relationship between the original fields and the prepotentials is rather different for the two systems - they involve only first-order derivatives for the lower spin multiplet, but also second-order derivatives for the higher spin multiplet -, the final formulas giving the supersymmetry transformation rules are remarkably similar and simple. 

In both cases, the transformation of the fermionic prepotential involves the combination $M_{ij} \equiv \Phi_{ij}-\gamma_{5}P_{ij}$ (or $M_i \equiv A^2_{i}-\gamma_{5}A^1_{i}$) of the two bosonic prepotentials that transforms under duality as
$$ \delta_{\text{dual}} M_{ij} = - \ \alpha \ \gamma_5 M_{ij} $$
(or $ \delta_{\text{dual}} M_{i} = - \ \alpha \ \gamma_5 M_{i} $).  At the same time,  the transformation of the first bosonic prepotential $\Phi_{ij}$ (or $A^1_i$) involves the function $\chi_{ij}$ (or $\psi_i$) of the fermionic prepotential that has identical gauge transformation properties as $\Phi_{ij}$ (or $A^1_i$), while the transformation of the second prepotential involves $-\gamma_5$ times it.


\section{Action}
\label{FormOfSusyAction}
\setcounter{equation}{0}

The action for the combined spin-$(2,5/2)$ system can be written in terms of the prepotentials. One finds by direct substitution,
\be
S[Z^a_{ij}, \Sigma_{ij}] = \int dt \left[ \int d^3 \, x \left( -2 \ep^{ab} D^{ij}_a \dot{Z} _{bij}  + \Theta^{A}  \dot{\psi}_{A}\right) - H \right], \label{FinalAction}
\ee
with 
\be
H =\int d^3x \left[ \left(4 R^a_{ij} R^{bij} - \frac{3}{2} R^a R^b \right) \de_{ab} + \mathcal{H} \right]. \label{HFinal}
\ee
Here, $D_a^{\; ij} \equiv D^{\; ij}[Z_a]$ (respectively $R_a^{\; ij} \equiv R^{\; ij}[Z_a]$) is the co-Cotton tensor (respectively, the Ricci tensor) constructed out of the prepotential $Z_{aij}$ \cite{Bunster:2012km}, whereas $\Theta^{A}(\Xi(\Sigma), \psi(\Sigma))$ and $\mathcal{H}(\Xi(\Sigma),\psi(\Sigma))$ are the functions of the second and third order derivatives of the prepotential $\Sigma_{ij}$ obtained by merely substituting (\ref{XiSigma1}) and (\ref{psiSigma1}) in (\ref{IV_theta_00}), (\ref{IV_theta_kl}) and (\ref{IV_hamiltonian}).

The action (\ref{FinalAction}) is manifestly separately invariant under duality rotations of the bosonic prepotentials $Z^a_{ij}$
\begin{equation}
\delta_{\text{dual}} Z^1_{ij} = \alpha Z^2_{ij}, \; \; \; \; \;  \delta_{\text{dual}} Z^2_{ij} = - \alpha Z^1_{ij}
\end{equation}
and chirality rotations of the prepotential $\Sigma_{ij}$,
\begin{equation}
\delta_{\text{chiral}} \Sigma_{ij} = \lambda \gamma_5 \Sigma_{ij},
\end{equation}
which implies $\delta_{\text{chiral}} \psi_{ij} = \lambda \gamma_5 \psi_{ij}$.

As such, neither the duality rotations (with untransforming fermions) nor the chirality transformations (with untransforming bosons) commute with supersymmetry. One can, however,  extend the duality transformation to the fermionic prepotential in such a way that duality and supersymmetry commute.  This is done by combining the duality rotations of the bosonic superpotentials with a chirality transformation of the fermionic prepotential of amplitude $ - \alpha$, 
\begin{eqnarray}
\delta_{\text{dual}} \Sigma_{jk} &=&  - \alpha \ \gamma_5 \Sigma_{jk} .
\end{eqnarray}
This commutativity property, however, would not hold for extended supersymmetry where only one supersymmetry would commute with duality.


\section{Comments and Conclusions}
\label{Conclusions}
\setcounter{equation}{0}

IIn this paper, we have made manifest the duality invariance for the combined Einstein-spin$\frac52$ system, which is the multiplet of hypergravity.   As found in earlier analyses, one cannot simultaneously have manifest duality invariance, and manifest Lorentz invariance.
 As argued in \cite{Bunster:2012hm}, this might signal that duality symmetry is more fundamental than spacetime covariance, in the sense that one might derive the latter from the former.

We have also exhibited a great similarity between the spin-($2,5/2$) and spin-($1,3/2$) systems.  

We have considered only the free theory.  The introduction of interactions requires an infinite tower of bosonic and fermionic fields of increasing spins, for which the same prepotential analysis would be needed to make duality manifest.
It is tempting in this context to guess the pattern that would hold for higher spins.  On the basis of what we found, one expects bosonic higher spin fields of spin $s$ to be described by two totally symmetric prepotentials $\Phi_{i_1 \cdots i_s}$ and $P_{i_1 \cdots i_s}$ with gauge symmetries
\begin{equation} \delta \Phi_{i_1 \cdots i_s} = \partial_{(i_1} \xi_{i_2 \cdots i_s)} + \delta_{(i_1 i_2} \epsilon_{i_3 \cdots i_s)} 
\end{equation}
and 
\begin{equation} \delta P_{i_1 \cdots i_s} = \partial_{(i_1} \xi'_{i_2 \cdots i_s)} + \delta_{(i_1 i_2} \epsilon'_{i_3 \cdots i_s)} 
\end{equation}
while a fermionic higher spin field of spin $s = \frac12$ would be described by a tensor spinor $\Sigma_{i_1 \cdots i_s}$ with gauge symmetries
\begin{equation} \delta \Sigma_{i_1 \cdots i_s} = \partial_{(i_1} \mu_{i_2 \cdots i_s)} + \gamma_{(i_1} \eta_{i_2 \cdots i_s)} .
\end{equation}
Both sets of gauge transformations are reducible since $\delta \Phi_{i_1 \cdots i_s}= 0$ for $\xi_{i_2 \cdots i_s} = \delta_{(i_2 i_3}\psi_{i_4 \cdots i_s)}$ and $\epsilon_{i_3 \cdots i_s} = -\partial_{(i_3}\psi_{i_4 \cdots i_s)}$,  while $\delta \Sigma_{i_1 \cdots i_s} = 0$ for $\mu_{i_2 \cdots i_s} = \gamma_{(i_2} \zeta_{i_3 \cdots i_s)}$ and $\eta_{i_2 \cdots i_s} = - \partial_{(i_2} \zeta_{i_3 \cdots i_s)}$.  The supersymmetry transformation would involve naturally the combination $\Phi_{i_1 \cdots i_s} - \gamma_5 P_{i_1 \cdots i_s}$ of the bosonic prepotentials as well as the function $\chi_{i_1 \cdots i_s}$ of the prepotential $\Sigma_{i_1 \cdots i_s}$ that has identical gauge symmetries as the bosonic prepotentials.

Although we leave the investigation of higher spins to future work, it is interesting to verify already here that the above ansatz satisfies a simple counting consistency requirement.  Both bosonic prepotentials would involve $\frac{k^2 + 3k +2}{2}$ (number of components of each prepotential) minus $ \frac{k^2 + k}{2}$ (number of components of $\xi_{i_1 \cdots i_{s-1}}$) minus $\frac{k^2 -k}{2}$ (number of components of $\epsilon_{i_1 \cdots i_{s-2}}$)  plus $\frac{k^2 - 3k +2}{2}$ (number of reducibility relations given by the number of components of $\psi_{i_1 \cdots i_{s-3}}$) $=2$ independent physical components, making a total of $4$ as it should if it is to  describe the two independent physical helicities of the massless spin-$s$ field.  Similarly, the fermionic prepotential $\Sigma_{i_1 \cdots i_s}$ would involve $4$ times [$\frac{k^2 + 3k +2}{2}$ (number of components of the prepotential) minus $2 \times \frac{k^2 + k}{2}$ (number of components of the  gauge parameters $\mu_{i_1 \cdots i_{s-1}}$ and $\eta_{i_1 \cdots i_{s-1}}$) plus $\frac{k^2 -k}{2}$ (number of components of $\zeta_{i_1 \cdots i_{s-2}}$)], which is again equal to $4$ as it must. Note that one may remove the redundancy in the gauge parameters by imposing the conditions $\delta^{i_1 i_2} \xi_{i_1 \cdots i_{s-1}} = 0$ on the bosonic gauge parameters $\xi_{i_1 \cdots i_{s-1}}$ and $\gamma^{i_1} \mu_{i_1 \cdots i_{s-1}} = 0$ on the fermionic gauge parameters $\mu_{i_1 \cdots i_{s-1}}$.

Again, the gauge symmetries of the  prepotentials take exactly the same form as the gauge symmetries of the corresponding free conformal field theory in 4 dimensions \cite{Fradkin:1985am}.

The supersymmetry parameter is a standard spin-1/2 spinor  $\epsilon$. With the inclusion of higher spin-fields, one expects supersymmetries of higher spins to also emerge \cite{Hietarinta:1975fu}, including the hypersymmetry of \cite{Aragone:1979hx}, the infinitesimal parameter of which is a vector-spinor $\epsilon_\mu$. 

Finally, our method for exhibiting duality through the introduction of prepotentials by solving the constraints should be contrasted with the approach in terms of the transverse degrees of freedom considered in \cite{Deser:2004xt}.  It would also be interesting to extend the analysis to include a cosmological constant \cite{Julia:2005ze}.


\section*{Acknowledgments} 
C.B. and M.H.  thank  the Alexander von Humboldt Foundation for Humboldt Research Awards.  A.L. is Research Fellow at the Belgian F.R.S.-FNRS. The work of M.H.,  S.H. and A. L. is partially supported by the ERC through the ``SyDuGraM" Advanced Grant, by IISN - Belgium (conventions 4.4511.06 and 4.4514.08) and by the ``Communaut\'e Fran\c{c}aise de Belgique" through the ARC program.  The Centro de Estudios Cient\'{\i}ficos (CECS) is funded by the Chilean Government through the Centers of Excellence Base Financing Program of Conicyt.   

\break

\noindent
{\bf \Large{Appendices}}

\appendix

\section{Technical derivations}
\label{Technical}
\setcounter{equation}{0}

\subsection{Bosonic prepotential $Z_{2}^{mn}=\phi^{mn}$}

The supersymmetry transformation rule of the graviton is

\bea
\delta h_{ij}&=&8i\bar{\epsilon}\psi_{ij}=8i\bar{\epsilon}\left[\delta_{ij}\Delta \Sigma-\delta_{ij}\partial^{a}\partial^{b}\Sigma_{ab}-2\Delta \Sigma_{ij}\right.\nonumber\\
&-&\left.\gamma_{ia}\partial^{a}\partial^{b}\Sigma_{jb}-\gamma_{ja}\partial^{a}\partial^{b}\Sigma_{ib}+\gamma_{ja}\Delta \Sigma^{a}_{\ i}+\gamma_{ia}\Delta \Sigma^{a}_{\ j}\right.\nonumber\\
&+&\left.\partial_{i}\epsilon_{j}+\partial_{j}\epsilon_{i}\right]\label{susyh}
\eea
On the other hand
\begin{equation}
\delta h_{ij}=\partial^{r}\epsilon_{irs}\delta\phi^{s}_{\ j}+\partial^{r}\epsilon_{jrs}\delta\phi^{s}_{\ i}+\partial_{i}\delta v_{j}+\partial_{j}\delta v_{i}\end{equation}
In order to compare these two expressions and find out the form of the supersymmetry transformation of the graviton prepotential, $\delta\phi_{ij}$, it is useful to recast (\ref{susyh}) in the form 
\begin{equation}
\delta h_{ij}=-8i\bar{\epsilon}\left[\partial^{r}\epsilon_{irs}\chi^{s}_{\ j}+\partial^{r}\epsilon_{jrs}\chi^{s}_{\ i}+\partial_{i}\eta_{j}+\partial_{j}\eta_{i}\right]
\end{equation}
This can be accomplished by setting 
\be
\chi_{js}=\frac{1}{2}\left[\epsilon_{jmn}\partial^{m}\Sigma^{n}_{\ s}+\epsilon_{smn}\partial^{m}\Sigma^{n}_{\ j}-\epsilon_{lmn}\gamma_{j}^{\ l}\partial^{m}\Sigma_{s}^{\ n}-\epsilon_{lmn}\gamma_{s}^{\ l}\partial^{m}\Sigma_{j}^{\ n}\right]\label{chi}
\ee
and
\be
\epsilon_{j}=\eta_{j}-\frac{1}{2}\partial_{j}\Sigma+\frac{3}{2}\partial^{r}\Sigma_{jr}+\frac{1}{2}\partial^{l}\gamma_{jl}\Sigma+\frac{1}{2}\gamma^{al}\partial_{l}\Sigma_{ja}-\frac{1}{2}\partial_{r}\gamma_{ja}\Sigma^{ra}
\ee
Thus one concludes that, up to a gauge transformation,
\begin{eqnarray}
&&\delta\Phi_{ij}=-8i\bar{\epsilon}\chi_{ij}\\
&&\hspace{-1cm} =-8i\bar{\epsilon}\frac{1}{2}(\epsilon_{jmn}\partial^{m}\Sigma_{i}^{\ n}+\epsilon_{imn}\partial^{m}\Sigma_{j}^{\ n}-\epsilon_{lmn}\gamma_{j}^{\ l}\partial^{m}\Sigma_{i}^{\ n}-\epsilon_{lmn}\gamma_{i}^{\ l}\partial^{m}\Sigma_{j}^{\ n})
\hspace{0.5cm} \end{eqnarray}
and
\begin{eqnarray}
&&\delta v_{i}=-8i\bar{\epsilon}\eta_{i} \\
&&\hspace{-1cm}
=-8i\bar{\epsilon}(\epsilon_{i}+\frac{1}{2}\partial_{i}\Sigma-\frac{3}{2}\partial^{n}\Sigma_{in}-\frac{1}{2}\partial^{r}\gamma_{ir}\Sigma+\frac{1}{2}\gamma_{is}\partial_{r}\Sigma^{sr}-\frac{1}{2}\partial^{r}\gamma_{sr}\Sigma_{i}^{\ s}) 
\hspace{0.5cm}\end{eqnarray}

Note that the field $\chi_{ij}$ defined by (\ref{chi}) transforms as 
\begin{equation}
\delta \chi_{ij} = \partial_i \alpha_j + \partial_j \alpha_i  + \delta_{ij} \beta
\end{equation}
with
\begin{equation}
\alpha_i = \frac12 \left(\epsilon_{imn} - \epsilon_{lmn} \gamma_i^{\; \, l} \right)\partial^m \mu^n + \tilde{\gamma}_5 \eta_i
\end{equation}
and
\begin{equation}
\beta = \epsilon_{lmn} \gamma^l \partial^m \eta^n
\end{equation}
under the gauge transformations (\ref{TransSigma}) of the prepotential $\Sigma_{ij}$.

\subsection{Bosonic prepotential $Z_{1}^{mn}=P^{mn}$}

Using $\delta h_{ij}=8i\bar{\epsilon}\psi_{ij}$ and the equation of motion of the hypergraviton
\be
\gamma_{\rho}\partial^{\rho}\psi_{\mu\nu}-\partial_{\mu}\gamma^{\rho}\psi_{\rho\nu}-\partial_{\nu}\gamma^{\rho}\psi_{\rho\mu}=0
\ee
one may write the supersymmetry transformation of the extrinsic curvature $K_{ij}=-\frac{1}{2}(\partial_{0}h_{ij}-\partial_{i}h_{0j}-\partial_{j}h_{0i})$ as
\be
\delta K_{ij}= -2 i \bar{\epsilon}\epsilon^{mkl} \gamma_5 \gamma_{kl}  \left[\partial_m \psi_{ij} - \partial_i \psi_{mj} - \partial_j \psi_{mi} \right]. \label{exc}
\ee
Substituting the expression of $\psi_{ij}$ in terms of the fermionic prepotential $\Sigma_{ij}$, one then 
finds
\begin{eqnarray}
\delta K_{ij} &=&
- \ 4i \bar{\epsilon} \gamma^0 \left[
\delta_{ij}\gamma_k \partial^k \left(\Delta \Sigma \ - \ \partial^m \partial^n \Sigma_{mn} \right)
\ - \ 2 \ \gamma_k \partial^k \Delta _{ij}
\right.
\nonumber \\ &&
\left. \qquad \qquad
\ + \ 2 \ \gamma_{m} \partial_i\partial^m \partial^n \Sigma_{nj} 
\ + \ 2 \ \gamma_{m} \partial_j\partial^m \partial^n \Sigma_{ni}
\right.
\nonumber \\ &&
\left. \qquad \qquad
\ + \ \gamma_{kim} \partial^k\Delta \Sigma^{m}_{\phantom{m}j}
\ + \ \gamma_{kjm} \partial^k\Delta \Sigma^{m}_{\phantom{m}i}
\right.
\nonumber \\ &&
\left. \qquad \qquad
\ + \ \gamma_{kjm} \partial^m \partial_n \partial_i \Sigma^{nk}
\ + \ \gamma_{kim} \partial^m \partial_n \partial_j \Sigma^{nk}
\right] .
\end{eqnarray}
As a corrolary, we get :
\begin{eqnarray}
\delta K &=&
- \ 4i \bar{\epsilon} \gamma^0 \left[ \gamma_k \partial^k \Delta \Sigma
\ + \ \gamma_{m} \partial^i\partial^m \partial^n \Sigma_{ni}
\right] .
\end{eqnarray}
and thus 
\begin{eqnarray}
\delta \pi_{ij} &=& - \ \delta K_{ij} \ + \ \delta_{ij} \delta K
\nonumber \\ &=&
4i \bar{\epsilon} \gamma^0 \left[
\ - \ 2 \ \delta_{ij}\gamma_k \partial^k  \partial^m \partial^n \Sigma_{mn}
\ - \ 2 \ \gamma_k \partial^k \Delta \Sigma_{ij}
\ + \ 2 \ \gamma_{m} \partial_i\partial^m \partial^n \Sigma_{nj}
\right.
\nonumber \\ &&
\left. \qquad \qquad
\ + \ 2 \ \gamma_{m} \partial_j\partial^m \partial^n \Sigma_{ni}
\ - \ \epsilon_{ikm} \gamma^0 \gamma_5 \partial^k\Delta \Sigma^{m}_{\phantom{m}j}
\ - \ \epsilon_{jkm} \gamma^0 \gamma_5 \partial^k\Delta \Sigma^{m}_{\phantom{m}i}
\right.
\nonumber \\ &&
\left. \qquad \qquad
\ + \ \epsilon_{ikm} \gamma^0\gamma_5 \partial^k \partial_n \partial_j \Sigma^{nm}
\ + \ \epsilon_{jkm} \gamma^0 \gamma_5 \partial^k \partial_n \partial_i \Sigma^{nm}
\right] \label{DeltaPi007}
\end{eqnarray}

Now, knowing the supersymmetry transformation $\delta\Phi_{ij}=-8i\bar{\epsilon}\chi_{ij}$ of the prepotential $\Phi_{ij}$ and comparing with the spin-1-spin-3/2 system, it is natural to guess that the supersymmetry transformation of the other bosonic prepotential $P_{ij}$ is simply 
\be
\delta P_{ij}=8i\bar{\epsilon}\gamma_5\chi_{ij} \label{DeltaChi007}
\ee

To prove this claim, it suffices to compute $\delta \pi_{ij}$ from its definition in terms ogf $P_{ij}$ and (\ref{DeltaChi007}),  and verify that the resulting expression coincides with (\ref{DeltaPi007}).  The computation is direct.  One finds
\begin{eqnarray}
\delta \pi_{ij} &=&
- \ 8i\bar{\epsilon}\gamma_5 \ \epsilon_{ikl} \epsilon_{jmn} \partial^k \partial^m \chi^{ln}
\nonumber \\ &=&
- \ 4 i\bar{\epsilon}\gamma_5 \ \left[ \epsilon_{ikl} \epsilon_{jmn}\epsilon^{lrs} \partial^k \partial^m \partial_r \Sigma_s^{\phantom{s}n}
\ + \ \epsilon_{ikl} \epsilon_{jmn}\epsilon^{nrs} \partial^k \partial^m \partial_r \Sigma_s^{\phantom{s}l}
\right.
\nonumber \\ && 
\left. \qquad \ \
\ - \ \epsilon_{ikl} \epsilon_{jmn} \epsilon_{prs} \gamma^{np} \partial^k \partial^m \partial^r \Sigma^{sl}
\ - \ \epsilon_{ikl} \epsilon_{jmn}\epsilon_{prs} \gamma^{lp} \partial^k \partial^m \partial^r \Sigma^{sn}
\right] \qquad \qquad
\nonumber \\ &=&
- \ 4i\bar{\epsilon}\gamma_5 \ \left[ 
 \epsilon_{jmn} \partial^k \partial^m \partial_i \Sigma_k^{\phantom{s}n}
\ + \ \epsilon_{ikl}  \partial^k \partial^m \partial_j \Sigma_m^{\phantom{m}l}
\right.
\nonumber \\ && 
\left. \qquad \ \
\ - \ \epsilon_{jmn}  \partial^m \Delta \Sigma_i^{\phantom{s}n}
\ - \ \epsilon_{ikl} \partial^k \Delta \Sigma_j^{\phantom{s}l}
\right.
\nonumber \\ && 
\left. \qquad \ \
\ - \ 2 \ \delta_{ij} \epsilon_{prs} \gamma^{np} \Delta \partial^r \Sigma^{s}_{\phantom{s}n}
\ + \ 2 \ \delta_{ij} \epsilon_{prs} \gamma^{np} \partial_n \partial_m \partial^r \Sigma^{sm}
\right.
\nonumber \\ && 
\left. \qquad \ \
\ - \ \epsilon_{prs} \gamma^{np} \partial_n \partial_i \partial^r \Sigma^{s}_{\phantom{s}j}
\ - \ \epsilon^{prs} \gamma_{jp} \partial^n \partial_i \partial_r \Sigma_{sn}
\right.
\nonumber \\ && 
\left. \qquad \ \
\ - \ \epsilon^{prs} \gamma_{ip} \partial_j \partial^m \partial_r \Sigma_{sm}
\ - \ \epsilon_{prs} \gamma^{mp} \partial_j \partial_m \partial^r \Sigma^{s}_{\phantom{s}i}
\right.
\nonumber \\ && 
\left. \qquad \ \
\ + \ \epsilon^{prs} \gamma_{ip} \Delta \partial_r \Sigma_{sj}
\ + \ \epsilon^{prs} \gamma_{jp} \Delta \partial_r \Sigma_{si}
\right.
\nonumber \\ && 
\left. \qquad \ \
\ + \ \epsilon_{prs} \gamma^{np} \partial_j \partial_i \partial^r \Sigma^{s}_{\phantom{s}n}
\ + \ \epsilon_{prs} \gamma^{np} \partial_j \partial_i \partial^r \Sigma^{s}_{\phantom{s}n}
\right] . \qquad \qquad
\end{eqnarray}
Using the identity $\gamma^{np} = \gamma^0 \gamma_5 \epsilon^{npq} \gamma_q$, this expression becomes
\begin{eqnarray}
\delta \pi_{ij} &=&
- \ 4i \bar{\epsilon}\gamma^0 \ \left[ 
\ - \ \epsilon_{jmn}\gamma^0 \gamma_5 \partial^k \partial^m \partial_i \Sigma_k^{\phantom{s}n}
\ - \ \epsilon_{ikl} \gamma^0 \gamma_5 \partial^k \partial^m \partial_j \Sigma_m^{\phantom{m}l}
\right.
\nonumber \\ && 
\left. \qquad \ \
\ + \ \epsilon_{jmn} \gamma^0 \gamma_5 \partial^m \Delta \Sigma_i^{\phantom{s}n}
\ + \ \epsilon_{ikl} \gamma^0 \gamma_5\partial^k \Delta \Sigma_j^{\phantom{s}l}
\right.
\nonumber \\ && 
\left. \qquad \ \
\ - \ 2 \ \delta_{ij} \gamma_r  \partial^r \left(\Delta \Sigma
\ - \ \partial_n \partial_m\Sigma^{nm}\right)
\right.
\nonumber \\ && 
\left. \qquad \ \
\ - \ 2 \ \gamma_r\partial_j \partial^m \partial^r \Sigma_{im}
\ - \ 2 \ \gamma_r\partial^n \partial_i \partial^r \Sigma_{nj}
\right.
\nonumber \\ && 
\left. \qquad \ \
\ + \ 2 \ \gamma_r \Delta \partial^r \Sigma_{ij}
\ + \ 2 \  \gamma_r\partial_j \partial_i \partial^r \Sigma
\right] . \qquad \qquad
\end{eqnarray}

The term $2\gamma_r\partial_j \partial_i \partial^r \Sigma - 2 \delta_{ij} \gamma_r  \partial^r \Delta \Sigma$ is a gauge transformation of $\pi_{ij}$, and can be removed, giving :
\begin{eqnarray}
\delta \pi_{ij} &=&
- \ 4i\bar{\epsilon}\gamma^0 \ \left[ 
\ - \ \epsilon_{jmn}\gamma^0 \gamma_5 \partial^k \partial^m \partial_i \Sigma_k^{\phantom{s}n}
\ - \ \epsilon_{ikl} \gamma^0 \gamma_5 \partial^k \partial^m \partial_j \Sigma_m^{\phantom{m}l}
\right.
\nonumber \\ && 
\left. \qquad \ \
\ + \ \epsilon_{jmn} \gamma^0 \gamma_5 \partial^m \Delta \Sigma_i^{\phantom{s}n}
\ + \ \epsilon_{ikl} \gamma^0 \gamma_5\partial^k \Delta \Sigma_j^{\phantom{s}l}
\right.
\nonumber \\ && 
\left. \qquad \ \
\ + \ 2 \ \delta_{ij} \gamma_r  \partial^r \partial_n \partial_m\Sigma^{nm}
\ + \ 2 \ \gamma_r \Delta \partial^r \Sigma_{ij}
\right.
\nonumber \\ && 
\left. \qquad \ \
\ - \ 2 \ \gamma_r\partial_j \partial^m \partial^r \Sigma_{im}
\ - \ 2 \ \gamma_r\partial^n \partial_i \partial^r \Sigma_{nj}
\right] , \qquad \qquad
\end{eqnarray}
which is exactly (\ref{DeltaPi007}).  This proves the correctness of (\ref{DeltaChi007}).

\subsection{Supersymmetric variation of the fermionic prepotential $\Sigma_{ij}$}

One can rewrite the supersymmetry transformation for $\psi_{\mu\nu}$ as
\bea
\delta\psi_{ij}&=&(\partial_{i}h_{j\rho}+\partial_{j}h_{i\rho})\gamma^{\rho}\epsilon-2\partial_{\rho}h_{ij}\gamma^{\rho}\epsilon+(\epsilon_{i\lambda\sigma\rho}\partial^{\lambda}h_{j}^{\ \sigma}+\epsilon_{j\lambda\sigma\rho}\partial^{\lambda}h_{i}^{\ \sigma})\gamma^{\rho}\gamma_{5}\epsilon\nonumber\\
&=&\delta_{\pi}\psi_{ij}+\delta_{h}\psi_{ij}\label{psi}
\eea
where
\bea
&&\delta_{\pi}\psi_{ij}=(\partial_{i}h_{j0}+\partial_{j}h_{i0})\gamma^{0}\epsilon-2\partial_{0}h_{ij}\gamma^{0}\epsilon\nonumber\\
&&+(\epsilon_{i0km}\partial^{0}h_{j}^{\ k}+\epsilon_{j0km}\partial^{0}h_{i}^{\ k}-\epsilon_{i0km}\partial^{k}h_{j}^{\ 0}-\epsilon_{j0km}\partial^{k}h_{i}^{\ 0})\gamma^{m}\gamma_{5}\epsilon\nonumber\\
&&\delta_{h}\psi_{ij}=(\partial_{i}h_{jk}+\partial_{j}h_{ik})\gamma^{k}\epsilon-2\partial_{k}h_{ij}\gamma^{k}\epsilon+(\epsilon_{ikl0}\partial^{k}h_{j}^{\ l}+\epsilon_{jkl0}\partial^{k}h_{i}^{\ l})\gamma^{0}\gamma_{5}\epsilon\nonumber\\
&&
\eea
We compute $\delta_{\pi}\psi_{ij}$ and $\delta_{h}\psi_{ij}$ separately.

\subsubsection{Term depending on the prepotential $Z^{1}_{mn}=P_{mn}$}

Let us first focus  on the term involving the extrinsic curvature. Adding to it the gauge transformation $\delta\psi_{ij}=(\partial_{i}h_{j0}+\partial_{j}h_{i0})\gamma^{0}\epsilon-(\epsilon_{i0km}\partial_{j}h^{k0}+\epsilon_{j0km}\partial_{i}h^{k0})\gamma^{m}\gamma_{5}\epsilon$ yields:
\bea
&&\delta_{\pi}\psi_{ij}=2(\partial_{i}h_{j0}+\partial_{j}h_{i0}-\partial_{0}h_{ij})\gamma^{0}\epsilon\nonumber\\
&&+(\epsilon_{i0km}\partial^{0}h_{j}^{\ k}-\epsilon_{i0km}\partial^{k}h_{j}^{\ 0}-\epsilon_{i0km}\partial_{j}h^{k0}+\epsilon_{j0km}\partial^{0}h_{i}^{\ k}-\epsilon_{j0km}\partial^{k}h_{i}^{\ 0}-\epsilon_{j0km}\partial_{i}h^{k0})\gamma^{m}\gamma_{5}\epsilon\nonumber\\
&&=4K_{ij}\gamma^{0}\epsilon-2\epsilon_{0ikm}K_{j}^{\ k}\gamma^{m}\gamma_{5}\epsilon-2\epsilon_{0jkm}K_{i}^{\ k}\gamma^{m}\gamma_{5}\epsilon\nonumber\\
&&=4(-\pi_{ij}+\frac{\pi}{2}\delta_{ij})\gamma^{0}\epsilon+2\epsilon_{0ikm}\pi_{j}^{\ k}\gamma^{m}\gamma_{5}\epsilon+2\epsilon_{0jkm}\pi_{i}^{\ k}\gamma^{m}\gamma_{5}\epsilon
\eea
Using,
\bea
&&-4\pi_{ij}+2\pi\delta_{ij}=-4\epsilon_{iab}\epsilon_{jcd}\partial^{a}\partial^{c}P^{bd}+2\delta_{ij}\epsilon_{mab}\epsilon^{m}_{\ cd}\partial^{a}\partial^{c}P^{bd}\nonumber\\
&&=-4\epsilon_{iab}\epsilon_{jcd}\partial^{a}\partial^{c}P^{bd}+\epsilon_{ixy}\epsilon_{jxy}\epsilon_{mab}\epsilon^{m}_{\ cd}\partial^{a}\partial^{c}P^{bd}\nonumber\\
&&=-2\epsilon_{iab}\epsilon_{jcd}\partial^{a}\partial^{c}P^{bd}-\epsilon_{jmb}\epsilon^{m}_{\ cd}\partial_{i}\partial^{c}P^{bd}-\epsilon_{imb}\epsilon^{m}_{\ cd}\partial_{j}\partial^{c}P^{bd}+\epsilon_{jma}\epsilon^{m}_{\ cd}\partial^{a}\partial^{c}P_{i}^{\ d}\nonumber\\
&&+\epsilon_{ima}\epsilon^{m}_{\ cd}\partial^{a}\partial^{c}P_{j}^{\ d}
\eea
one finds
\bea
&&\delta_{\pi}\psi_{ij}=-\left[2\epsilon_{iab}\epsilon_{jcd}\partial^{a}\partial^{c}P^{bd}-\epsilon_{jma}\epsilon^{m}_{\ cd}\partial^{a}\partial^{c}P_{i}^{\ d}-\epsilon_{ima}\epsilon^{m}_{\ cd}\partial^{a}\partial^{c}P_{j}^{\ d}\right]\gamma^{0}\epsilon\nonumber\\
&&+2\left[\epsilon_{0jnm}\epsilon_{iab}\epsilon^{n}_{\ cd}\partial^{a}\partial^{c}P^{bd}+\epsilon_{0inm}\epsilon_{jab}\epsilon^{n}_{\ cd}\partial^{a}\partial^{c}P^{bd}\right]\gamma^{m}\gamma_{5}\epsilon\nonumber\\
&&-\epsilon_{jmb}\epsilon^{m}_{\ cd}\partial_{i}\partial^{c}P^{bd}\gamma^{0}\epsilon-\epsilon_{imb}\epsilon^{m}_{\ cd}\partial_{j}\partial^{c}P^{bd}\gamma^{0}\epsilon\label{psi-pi}
\eea
and by dropping again a gauge transformation for $\psi_{ij}$ the following supersymmetry transformation rule may be derived for $\chi_{ij}$:
\be
\delta\chi_{j}^{\ b}=2\partial_{m}P_{j}^{\ b}\gamma^{m}\gamma_{5}\epsilon-\epsilon_{jcd}\partial^{c}P^{bd}\gamma^{0}\epsilon-\epsilon^{bcd}\partial_{c}P_{jd}\gamma^{0}\epsilon\label{chi1}
\ee

\subsubsection{Term depending on the $Z^{2}_{mn}=\phi_{mn}$ prepotential}

Turn now to $\delta_{h}\psi_{ij}$. Expressing it in terms of the prepotential $\phi_{mn}$ gives
\bea
&&\delta_{h}\psi_{ij}=(\partial_{i}h_{jk}+\partial_{j}h_{ik})\gamma^{k}\epsilon-2\partial_{k}h_{ij}\gamma^{k}\epsilon+(\epsilon_{ikl0}\partial^{k}h_{j}^{\ l}+\epsilon_{jkl0}\partial^{k}h_{i}^{\ l})\gamma^{0}\gamma_{5}\epsilon\nonumber\\
&&=\partial_{i}(\partial^{l}\epsilon_{jlm}\phi^{m}_{\ k}+\partial^{l}\epsilon_{klm}\phi^{m}_{\ j})\gamma^{k}\epsilon+\partial_{j}(\partial^{l}\epsilon_{ilm}\phi^{m}_{\ k}+\partial^{l}\epsilon_{klm}\phi^{m}_{\ i})\gamma^{k}\epsilon\nonumber\\
&&-2\partial_{k}(\partial^{l}\epsilon_{ilm}\phi^{m}_{\ j}+\partial^{l}\phi^{m}_{\ i})\gamma^{k}\epsilon+\epsilon_{ikl0}\partial^{k}(\partial^{m}\epsilon_{jmn}\phi^{nl}+\partial^{m}\epsilon^{l}_{\ mn}\phi^{n}_{\ j})\gamma^{0}\gamma_{5}\epsilon\nonumber\\
&&+\epsilon_{jkl0}\partial^{k}(\partial^{m}\epsilon_{imn}\phi^{nl}+\partial^{m}\epsilon^{l}_{\ mn}\phi^{n}_{\ i})\gamma^{0}\gamma_{5}\epsilon
\eea
which, up to a gauge transformation, contributes to the supersymmetry transformation rule of $\chi_{ij}$ as follows:
\be
\delta\chi_{jm}=-2\partial_{k}\phi_{jm}\gamma^{k}\epsilon-(\partial^{n}\epsilon_{jnp}\phi^{p}_{\ m}+\partial^{n}\epsilon_{mnp}\phi^{p}_{\ j})\gamma^{0}\gamma_{5}\epsilon\label{chi2}
\ee
Finally, by adding up (\ref{chi1}) and (\ref{chi2}) the complete supersymmetry transformation rule for the fermionic prepotential $\chi_{ij}$ is obtained:
\be
\delta\chi_{ij}=-2\partial_{k}(\phi_{ij}+P_{ij}\gamma_{5})\gamma^{k}\epsilon+\epsilon_{icd}\partial^{c}(\phi^{d}_{\ j}\gamma_{5}-P_{j}^{\ d})\gamma^{0}\epsilon+\epsilon_{jcd}\partial^{c}(\phi^{d}_{\ i}\gamma_{5}-P_{i}^{\ d})\gamma^{0}\epsilon
\ee

Using the identity $\epsilon_{lxy}\gamma_{i}^{\ l}=(\delta_{ix}\gamma_{y}-\delta_{iy}\gamma_{x})\gamma_{0}\gamma_{5}$ in (\ref{chi}), one can see immediatly that this implies the following transformation rule for the prepotential $\Sigma_{ij}$,
\be
\delta \Sigma_{ij}=2\gamma_{5}\gamma_{0}(\phi_{ij}-\gamma_{5}P_{ij})\epsilon
\ee

\section{The spin-$(1,\frac32)$) multiplet}
\label{132Mult}
\setcounter{equation}{0}

\subsection{Spin 1}

The covariant action of the Maxwell field 
\begin{equation}
S_{1} [A_\mu]
=
- \ \frac{1}{4} \ \int \ d^4 x \ F_{\mu\nu} F^{\mu\nu}
 \label{action_lag_1}
\end{equation}
can be recast in a manifestly duality-invariant form by going to the first-order formalism and introducing a second vector potential $A^2_i$ through the resolution of Gauss' contraint \cite{Deser:1976iy}.  One finds, in duality covariant notations \cite{Deser:1997mz},
\begin{equation}
I = \frac{1}{2} \int dx^0 d^3x \left( \ep_{ab} \vec{B}^a \cdot \dot{\vec{A}}^b - \de_{ab} \vec{B}^a \cdot \vec{B}^b \right). \label{TwoPotential}
\end{equation}
Here, $\ep_{ab}$ is given by $\ep_{ab} = - \ep_{ba}$, $\ep_{12} = +1$ and 
$$
\vec{B} ^a  = \vec{\nabla} \times \vec{A}^a,
$$
with $A_i^1 \equiv A_i$.

The action (\ref{TwoPotential}) is invariant under rotations in the $(1,2)$ plane (``electric-magnetic duality rotations") ,
\be
\begin{pmatrix} \vec{A}^1 \\ \vec{A}^2  \end{pmatrix} \equiv \vec{\mathbf{A}} \; \;  \longrightarrow \; \; e^{\alpha \ep} \vec{\mathbf{A}} \label{E-MDuality}
\ee
because $\ep_{ab}$ and $\de_{ab}$ are invariant tensors.  In infinitesimal form, 
\begin{eqnarray}
\delta_{\text{dual}} A^1_k &=& \alpha \ A^2_k , \label{dual_1}
\\
\delta_{\text{dual}} A^2_k &=& - \ \alpha \ A^1_k . \label{dual_2}
\end{eqnarray}
The action (\ref{TwoPotential}) is also invariant under $U(1) \times U(1)$ gauge transformations,
$$
A^a_k \; \;  \longrightarrow \; \; A_k^a + \partial_k \Lambda^a.
$$

\subsection{Spin 3/2}

The covariant action of the spin $3/2$ is given by the expression :
\begin{eqnarray}
S_{3/2} &=& 
i \ \int \ d^4 x \ 
\bar{\psi}_{\mu} \gamma^{\mu \nu \rho} \partial_{\nu} \psi_{\rho} \label{action_lag_3_2}
\end{eqnarray}
which is invariant under the gauge transformation $\delta \psi_{\mu} = \partial_{\mu} \epsilon $.

This first-order action is already in canonical form, with $\psi_k$ being self-conjugate canonical variables and $\psi_0$ the Lagrange multiplier for the constraint
\begin{eqnarray}
0 &=&
\gamma^{kl} \partial_k \psi_l . \label{constraint_psi}
\end{eqnarray}
The general solution of the constraint (\ref{constraint_psi}) reads \cite{Bunster:2012jp}:
\begin{eqnarray}
\psi_k &=& 
- \ \frac{1}{2} \ \epsilon^{lmn} \gamma_l \gamma_k \partial_m \chi_n , \label{def_chi}
\end{eqnarray}
\noindent where $\chi_k$ is a vector-spinor, which is the prepotential for the spin-3/2 field. The ambiguity in $\chi_k$  is given by \cite{Bunster:2012jp},
\begin{eqnarray}
\delta_{\text{gauge}} \chi_k &=& \partial_k \eta \ + \ \gamma^0 \gamma_5 \gamma_k \epsilon ,
\end{eqnarray}
\noindent where $\eta$ and $\epsilon$ are arbitrary spinor fields.  As observed in \cite{Bunster:2012jp}, these are the same gauge symmetries as those of a conformal spin-3/2 field.

\subsection{Supersymmetry}

The supersymmetry transformations for the $(1, 3/2)$-multiplet  read 
\begin{eqnarray}
\delta_{\text{SUSY}} A_{\mu} &=& i \ \bar{\epsilon}\psi_{\mu}, \label{susy_A}
\\
\delta_{\text{SUSY}} \psi_{\mu} &=& 
\frac{1}{4} \ F_{\mu \nu} \gamma^{\nu} \epsilon 
\ + \ \frac{1}{4} \ \tilde{F}_{\mu \nu} \gamma_5  \gamma^{\nu} \epsilon
. \qquad \label{susy_psi0}
\end{eqnarray}
and are easily verified to leave the covariant action $S_1 \ + \ S_{3/2}$ invariant.

The supersymmetry transformations can be rewritten in terms of the prepotentials $(A^a_k, \chi_k)$. 
From the supersymmetry transformation of the photon field (\ref{susy_A}), one immediately deduces that :
\begin{eqnarray}
\delta_{\text{SUSY} }A^{1}_{k} &=& i \ \bar{\epsilon} \psi_k , \label{SusyPrepA10}
\end{eqnarray}
where $\psi_k$ is now to be viewed as the function (\ref{def_chi}) of the prepotential $\chi_k$.  Similarly, a direct computation shows that the supersymmetry transformation of the momentum $\Pi^k$ conjugate to $A_k \equiv A^1_k$ ( minus the original electric field) is 
\begin{eqnarray}
\delta_{\text{SUSY}} \Pi^k &=&
- \ \frac{i}{2} \ \epsilon^{klm} \bar{\epsilon} \gamma_5 \partial_l \psi_m
\ + \ \frac{i}{2} \  \bar{\epsilon}\gamma^0  \gamma^{klm} \partial_l \psi_m 
\nonumber \\ &=&
- \ i \ \epsilon^{klm} \bar{\epsilon} \gamma_5 \partial_l \psi_m .
\end{eqnarray}
Since $\Pi^k = \epsilon^{klm} \partial_l A^2_m$, one gets
\begin{equation}
\delta_{\text{SUSY}} A^2_k =
- \ i \ \bar{\epsilon} \gamma_5 \psi_k  \label{SusyPrepA20}
\end{equation}
(up to a gauge transformation that can be set to zero).
Finally, one easily derives from  (\ref{susy_psi0})
\begin{eqnarray}
\delta_{\text{SUSY}} \psi_k &=& 
- \ \frac{1}{4} \ W_k \gamma^{0} \epsilon 
\ + \ \frac{1}{4} \ \epsilon_{klm} W^m \gamma_5  \gamma^{l} \epsilon .
\end{eqnarray}
where we have defined $W_k \equiv \Pi_k \ - \ B_k \gamma_5$.  It follows that, again up to a gauge transformation of the prepotential that can be set to zero, 
\begin{eqnarray}
\delta_{\text{SUSY}}\chi^k &=&
\frac{1}{2} \ \left(A^2_k \ - \ A^1_k \gamma_5 \right) \gamma^0  \epsilon \nonumber \\
&=& \frac{1}{2} \ M_k \gamma^0  \epsilon \label{SusyPrepChi0}
\end{eqnarray}
with 
\begin{equation}
M_k \equiv A^2_k \ - \ A^1_k \gamma_5 \, .
\end{equation}

The transformations (\ref{SusyPrepA10}), (\ref{SusyPrepA20}) and (\ref{SusyPrepChi0}) are the searched-for supersymmetry transformations in terms of the prepotentials.

We close by noting that the duality rotations  (\ref{dual_1}) and (\ref{dual_2}) with $\delta \chi = 0$, and the chirality rotations
\begin{eqnarray}
\delta_{\text{chiral}} \chi_{k} &=&  \lambda \ \gamma_5 \chi_{k}  \label{chirality32}
\end{eqnarray}
with $\delta A^1_k = \delta A^2_k = 0$, separately leave the action invariant.  None of these transformations commutes with supersymmetry. However, the combined duality-chirality transformation (\ref{dual_1}), (\ref{dual_2}) and (\ref{chirality32}) with $\lambda = - \alpha$ commutes with supersymmetry. This is because $M_k$ transforms as
\begin{eqnarray}
\delta_{\text{dual}} M_k &=& - \ \alpha \ \gamma_5 M_k 
\end{eqnarray}
under duality.  In the case of extended supersymmetry,  duality acting only on the vector fields will not commute either with supersymmetry, but again it can be redefined to do so with one of the supersymmetries.

\break

\end{document}